\documentstyle[aps,amssymb,twocolumn]{revtex}

\begin{document}
\draft

\title{Generalized coherent states are unique Bell states
of quantum systems with Lie group symmetries}
\author{C. Brif, \cite{email1} A. Mann, \cite{email2}
and M. Revzen \cite{email3}}
\address{Department of Physics, Technion -- Israel Institute of 
Technology, Haifa 32000, Israel}
\maketitle

\begin{abstract}
We consider quantum systems, whose dynamical symmetry groups are 
semisimple Lie groups, which can be split or decay into two subsystems
of the same symmetry. We prove that the only states of such a system 
that factorize upon splitting are the generalized coherent states. 
Since Bell's inequality is never violated by the direct product state, 
when the system prepared in the generalized coherent state is split, 
no quantum correlations are created. 
Therefore, the generalized coherent states are the unique Bell states, 
i.e., the pure quantum states preserving the fundamental classical 
property of satisfying Bell's inequality upon splitting.
\end{abstract}


Entanglement is the main quantum mechanical feature that distinguishes 
between quantum and classical physics. A state of a quantum system,
consisting of two subsystems, is called entangled, if it cannot be
represented as a direct product of states of the subsystems. The
nonclassicality of such a situation is expressed mathematically by
the violation of Bell's inequality \cite{Bell,CHSH69}. 
More specifically, let us consider a quantum system $A$ (e.g., a mode 
of the quantized radiation field or a spin-$j$ particle), which is 
split or decays into two subsystems $B$ and $C$ \cite{comment1}. 
A pure quantum state that upon splitting will not violate Bell's 
inequality (i.e., a quantum state that possesses the basic attribute 
of classical physics) has been called the Bell state \cite{MRS92}.
It has been proven \cite{MRS92,Peres78,Gisin91} that a state that
factorizes upon splitting into a direct product cannot violate Bell's 
inequality, while a state that gives rise to an entangled (i.e., 
non-product) state can always, by a suitable choice of local 
apparatus, be made to violate the inequality.
Therefore, the Bell state of the system $A$ must factorize upon 
splitting into the direct product of states of the subsystems.

In 1966 Aharonov {\em et al.\/} \cite{AFLP66} showed that the 
only states of the single-mode quantized radiation field that 
factorize upon splitting are the Glauber coherent states (CS) 
\cite{Glauber}. Therefore, the Glauber CS are the unique Bell states 
of the quantized radiation field \cite{MRS92}. The Glauber CS are
known to have a number of classical characteristics. Actually, the CS
were discovered in 1926 by Schr\"{o}dinger \cite{Schro26} who looked 
for a wave packet with minimum possible dispersions that will 
preserve its form while moving along a classical trajectory. 
The Glauber CS are such wave packets for a quantum harmonic 
oscillator (which is the mathematical model of the single-mode 
quantized radiation field). One of the criteria used by Glauber
\cite{Glauber} was to choose the quantum state of the radiation 
field produced by a classically prescribed current. 
Also, the Glauber CS have the property of maximal coherence
(in accordance to their name), i.e., their normal-ordered
correlation functions of all orders factorize \cite{comment2}. 

Formally, the Glauber CS $|\alpha\rangle$ ($\alpha \in \Bbb{C}$) 
can be defined in two equivalent ways: (i) as the eigenstates of the 
boson annihilation operator $a$,
\begin{equation}
a |\alpha\rangle = \alpha |\alpha\rangle
\label{def1} 
\end{equation}
(this property actually implies another possible definition of 
$|\alpha\rangle$ as minimum-uncertainty states), and (ii) as 
the states produced by the action of the displacement operator
$D(\alpha) = \exp(\alpha a^{\dagger} - \alpha^{\ast} a)$ on 
the vacuum,
\begin{equation}
|\alpha\rangle = D(\alpha) |0\rangle .
\label{def2} 
\end{equation}
Various generalizations of the Glauber states to quantum systems
other than the harmonic oscillator have been proposed. The most
useful of them is the group-theoretic generalization of CS 
developed by Perelomov \cite{Per} and Gilmore \cite{Gil,Gil:rev}. 
The generalized CS for a system, possessing a dynamical symmetry 
Lie group, are produced by the action of group elements 
on a reference state in the Hilbert space. For the Weyl-Heisenberg
group $W$, which is the dynamical symmetry group of a quantum
harmonic oscillator, group elements are represented by the 
displacement operator $D(\alpha)$, the reference state is the
vacuum, and the general definition of the CS reduces to 
Eq.\ (\ref{def2}).

Under the action of group elements, the generalized CS transform 
among themselves. For the evolution operator which is an element of 
the dynamical symmetry group, the generalized CS will evolve along 
a classical trajectory in the phase space of the group parameters.
While the generalized CS possess this classical feature, not all
the properties of the Glauber states are preserved under the 
Gilmore-Perelomov generalization. For example, not every generalized 
coherent state is a minimum-uncertainty state. However, we will show 
in the present paper that the main classical attribute of the 
Glauber CS is preserved under the group-theoretic generalization. 
Specifically, we will consider quantum systems, whose dynamical 
symmetry groups are semisimple Lie groups, and prove that the 
only states of such a system that factorize upon splitting are 
the generalized CS. Therefore, the generalized CS are the unique 
Bell states of quantum systems with semisimple Lie group symmetries.
In order to illustrate this general result, we will consider the 
decay of a spin-$j$ particle, with SU(2) as the dynamical symmetry 
group.

The theory of the generalized CS is well developed and elaborated
(e.g., see Refs.\ \cite{Per,Gil,Gil:rev}). Here we recapitulate some 
basic results which are essential for our discussion. Let $G$ be an 
arbitrary Lie group and $\Gamma_{\Lambda}$ its unitary irreducible 
representation acting on the Hilbert space ${\cal H}_{\Lambda}$. By 
choosing a fixed normalized reference state $|\Psi_{0}\rangle \in 
{\cal H}_{\Lambda}$, one can define the system of states 
$\{ |\Psi_{g}\rangle \}$, 
\begin{equation}
|\Psi_{g}\rangle = T(g) |\Psi_{0}\rangle , \hspace{1cm} g \in G , 
\end{equation}
which is called the coherent-state system. 
The isotropy (or maximum-stability) subgroup $H \subset G$ consists 
of all the group elements $h$ that leave the reference state 
invariant up to a phase factor,
\begin{equation}
T(h) |\Psi_{0}\rangle = e^{i\phi(h)} |\Psi_{0}\rangle , 
\hspace{1cm} h \in H .               
\end{equation}
For every element $g \in G$, there is a unique decomposition of $g$ 
into a product of two group elements, one in $H$ and the other in the 
coset space $G/H$,
\begin{equation}
g = \Omega h , \hspace{1cm} g \in G, \;\; h \in H, \;\; 
\Omega \in G/H .     
\end{equation}
It is clear that group elements $g$ and $g'$ with different $h$ and 
$h'$ but with the same $\Omega$ produce CS which differ only by a 
phase factor: 
$|\Psi_{g}\rangle = e^{i\delta} |\Psi_{g'}\rangle$, where 
$\delta =\phi(h) -\phi(h')$. 
Therefore a coherent state $|\Omega\rangle \equiv 
|\Psi_{\Omega}\rangle$ is determined by a point $\Omega = \Omega(g)$ 
in the coset space $G/H$. This coset space is actually the phase 
space of the system. One can see that the choice of the reference 
state $|\Psi_{0}\rangle$ determines the structure of the 
coherent-state system and of the phase space.

In order to make our discussion more concrete, we will consider 
the case when $G$ is a semisimple Lie group \cite{comment:discrete}. 
The corresponding Lie algebra $\frak{G}$ has a Cartan subalgebra 
$\frak{H}$, and the set of non-zero roots is denoted by $\Delta$ 
\cite{BaRa86}. We use the standard Cartan-Weyl basis 
$\{ H_{i},E_{\alpha} \}$; for $H_{i} \in \frak{H}$ and 
$\alpha,\beta \in \Delta$, the commutation relations are 
\cite{BaRa86}
\begin{mathletters}
\label{comrel}
\begin{eqnarray}
& & [H_{i},H_{j}] = 0 , \\
& & [H_{i},E_{\alpha}] = \alpha(H_{i}) E_{\alpha} , \\
& & [E_{\alpha},E_{\beta}] = \left\{ 
  \begin{array}{l}
0 , \;\;\;\; {\rm if}\; \alpha + \beta \neq 0\; {\rm and}\;
\alpha + \beta \notin \Delta , \\
H_{\alpha} , \;\;\;\; {\rm if}\; \alpha + \beta = 0 , \\
N_{\alpha,\beta} E_{\alpha + \beta} , \;\;\;\; {\rm if}\;
\alpha + \beta \in \Delta .
  \end{array} \right.
\end{eqnarray}
\end{mathletters}
The operators $H_{i}$, which constitute the Cartan subalgebra 
$\frak{H}$, may be taken as diagonal in any unitary irreducible 
representation $\Gamma_{\Lambda}$, while $E_{\alpha}$ are the 
``shift operators.'' One can always choose the representation 
$\Gamma_{\Lambda}$ such that $H_{i}^{\dagger} = H_{i}$ and 
$E_{\alpha}^{\dagger} = E_{-\alpha}$. Every group element 
$g \in G$ can be written as the exponential of an anti-Hermitian 
complex linear combination of $H_{i}$ and $E_{\alpha}$. 

As an illustrative example, we consider the most 
elementary compact non-Abelian simple Lie group SU(2), which is 
the dynamical symmetry group of a spin-$j$ particle. The su(2) 
Lie algebra is spanned by the three operators 
$\{ J_{0} , J_{+} , J_{-} \}$,
\begin{equation}
[J_{0} , J_{\pm}] = \pm J_{\pm}, \hspace{1.0cm} 
[J_{+} , J_{-}] = 2 J_{0} \ .   
\end{equation}
The Casimir operator
$J^{2} = J_{0}^{2} + (J_{+}J_{-} + J_{-}J_{+})/2$
for any unitary irreducible representation is the identity operator 
times a number: $J^{2} = j(j+1) I$.
Thus an irreducible representation $\Gamma_{j}$ of SU(2) is 
determined by a single number $j$ that can be a non-negative integer 
or half-integer: $j=0,\frac{1}{2},1,\frac{3}{2},2,\ldots$. The 
representation Hilbert space ${\cal H}_{j}$ is spanned by the 
orthonormal basis $|j,m\rangle$ ($m=j,j-1,\dots,-j$). The
Hermitian operator $J_{0}$ is diagonal, $J_{0} |j,m\rangle =
m |j,m\rangle$, while $J_{+}$ and $J_{-}$ are the raising and 
lowering operators, respectively, $J_{\pm} |j,m\rangle =
[(j \mp m)(j \pm m + 1)]^{1/2} |j,m \pm 1\rangle$ and 
$J_{+}^{\dagger} = J_{-}$. 

It is physically sensible to choose the reference state 
$|\Psi_{0}\rangle$ to be the ground state of the system, which 
is mathematically an ``extremal state'' in the Hilbert space 
${\cal H}_{\Lambda}$ \cite{comment3}. This ``extremal state'' is 
the lowest-weight state $|\Lambda,-\Lambda\rangle$ \cite{comment4}, 
which is annihilated by all the lowering operators $E_{\alpha}$ 
with $\alpha < 0$,
\begin{equation}
E_{\alpha} |\Lambda,-\Lambda\rangle = 0 , \hspace{1.0cm} 
\alpha < 0 ,    \label{lower}
\end{equation}
and mapped into itself by all the diagonal operators $H_{i}$,
\begin{equation}
H_{i} |\Lambda,-\Lambda\rangle = \Lambda_{i} 
|\Lambda,-\Lambda\rangle , \hspace{1.0cm} H_{i} \in \frak{H} .
\label{diag}
\end{equation}
The last equation means that the Lie algebra of the 
maximum-stability subgroup $H$ is just the Cartan subalgebra 
$\frak{H}$. Therefore, elements $\Omega$ of the coset space $G/H$ 
are the generalized displacement operators of the form
\begin{equation}
\Omega = \exp\left[ \sum_{\alpha>0} ( \eta_{\alpha} E_{\alpha} 
- \eta_{\alpha}^{\ast} E_{-\alpha} ) \right] .
\end{equation}
Here the parameters $\eta_{\alpha}$ are complex numbers and
the summation $\sum_{\alpha>0}$ is restricted to those raising
operators $E_{\alpha}$ ($\alpha > 0$) which do not annihilate 
the reference state, $E_{\alpha} |\Lambda,-\Lambda\rangle \neq 0$.
We also use the Baker-Campbell-Hausdorff formula \cite{Helg78}, 
\begin{eqnarray*}
& & \exp\left[ \sum_{\alpha>0} ( \eta_{\alpha} E_{\alpha} 
- \eta_{\alpha}^{\ast} E_{-\alpha} ) \right]  \\
& & = \exp\left[ \sum_{\alpha>0} \tau_{\alpha} E_{\alpha} \right]
\exp\left[ \sum_{i} \gamma_{i} H_{i} \right]
\exp\left[ -\sum_{\alpha>0} \tau_{\alpha}^{\ast} 
E_{-\alpha} \right] . 
\end{eqnarray*}
The relation between $\tau_{\alpha}$, $\gamma_{i}$ and 
$\eta_{\alpha}$ can be derived from the matrix representation of
$G$; for details see Ref.\ \cite{Gil:rev}. 
The generalized CS $|\Lambda,\Omega\rangle$ are given by
\begin{equation}
|\Lambda,\Omega\rangle = \Omega |\Lambda,-\Lambda\rangle
= {\cal N} \exp\left[ \sum_{\alpha>0} \tau_{\alpha} 
E_{\alpha} \right] |\Lambda,-\Lambda\rangle ,
\label{gencs}
\end{equation}
where 
\begin{equation}
{\cal N} = \exp\left[ \sum_{i} \gamma_{i} \Lambda_{i} \right]
\label{nfactor}
\end{equation}
is the normalization factor.

In the case of the SU(2) group, the lowest-weight state is
$|j,-j\rangle$, which is annihilated by $J_{-}$. 
The maximum-stability subgroup $H$=U(1) consists of all group 
elements $h$ of the form $h=\exp(i\delta J_{3})$. 
The coset space is SU(2)/U(1) (the sphere), and 
the coherent state is specified by a unit vector
\begin{equation}
\bbox{n} = (\sin\theta\cos\varphi,\sin\theta\sin\varphi,\cos\theta).
\end{equation}
Then an element $\Omega$ of the coset space can be written as 
\begin{equation}
\Omega = \exp (\xi J_{+} - \xi^{\ast} J_{-}) ,  
\end{equation}
where $\xi = -(\theta/2) e^{-i\varphi}$. The SU(2) CS are 
given by 
\begin{equation}
|j,\zeta\rangle =  \Omega |j,-j\rangle
= (1+|\zeta|^{2})^{-j} \exp(\zeta J_{+}) |j,-j\rangle , 
\label{su2cs} 
\end{equation}
where $\zeta = (\xi/|\xi|)\tan |\xi| = - \tan (\theta/2) 
e^{-i\varphi}$. 

Now we consider a quantum system $A$, whose dynamical symmetry 
group is $G$, that is split or decays into two subsystems $B$ 
and $C$ of the same symmetry \cite{comment:sym}. The commutation 
relations (\ref{comrel}) are satisfied if and only if
\begin{mathletters}
\label{decay}
\begin{eqnarray}
& & E_{A \alpha} = E_{B \alpha} \otimes I_{C} + 
I_{B} \otimes E_{C \alpha} ,  \label{decay:a} \\
& & H_{A i} = H_{B i} \otimes I_{C} + I_{B} \otimes H_{C i} ,
\label{decay:b}
\end{eqnarray}
\end{mathletters}
where $\alpha \in \Delta$ and $H_{X i} \in {\frak H}$, $X = A,B,C$.
In the case of the SU(2) group, conditions (\ref{decay}) are
equivalent to the rule for the addition of angular momenta,
\begin{equation}
\bbox{J}_{A} =  \bbox{J}_{B} \otimes I_{C} + 
I_{B} \otimes \bbox{J}_{C} .
\end{equation}
The lowest-weight state (which usually is the ground state of 
the quantum system) must factorize:
\begin{equation}
|\Lambda_{A},-\Lambda_{A}\rangle_{A} =  
|\Lambda_{B},-\Lambda_{B}\rangle_{B} \otimes 
|\Lambda_{C},-\Lambda_{C}\rangle_{C} .
\label{vacdecay}
\end{equation}

We first prove that any generalized coherent state factorizes
upon splitting. Indeed, using Eqs.\ (\ref{decay:a}) and 
(\ref{vacdecay}), we obtain
\begin{eqnarray*}
& & |\Lambda_{A},\Omega(\eta)\rangle_{A} = 
\exp\left[ \sum_{\alpha>0} ( \eta_{\alpha} E_{A \alpha} 
- \eta_{\alpha}^{\ast} E_{A \alpha}^{\dagger} ) \right]
|\Lambda_{A},-\Lambda_{A}\rangle_{A} \\
& & = \exp\left[ \sum_{\alpha>0} ( \eta_{\alpha} E_{B \alpha} 
- \eta_{\alpha}^{\ast} E_{B \alpha}^{\dagger} ) \right]
|\Lambda_{B},-\Lambda_{B}\rangle_{B} \\
& & \otimes \exp\left[ \sum_{\alpha>0} ( \eta_{\alpha} E_{C \alpha} 
- \eta_{\alpha}^{\ast} E_{C \alpha}^{\dagger} ) \right]
|\Lambda_{C},-\Lambda_{C}\rangle_{C} ,
\end{eqnarray*}
where we also use the basic property of the exponential function:
$\exp(x+y) = \exp(x) \exp(y)$, if $[x,y] = 0$ (of course, operators
describing different systems are commuting). Then we get
\begin{equation}
|\Lambda_{A},\Omega(\eta)\rangle_{A} = 
|\Lambda_{B},\Omega(\eta)\rangle_{B} 
\otimes |\Lambda_{C},\Omega(\eta)\rangle_{C} .
\end{equation}
We see that the coherent amplitudes $\eta_{\alpha}$ are the
same for the system $A$ and for the subsystems $B$ and $C$.
In particular, we have for the spin \cite{comment:spin}:
\begin{equation}
|j_{A},\zeta\rangle_{A} = 
|j_{B},\zeta\rangle_{B} \otimes |j_{C},\zeta\rangle_{C} ,
\end{equation}
i.e., $\zeta_{A} = \zeta_{B} = \zeta_{C}$. Note that the 
situation is rather different for the Glauber coherent state 
$|\alpha\rangle_{A}$ of the radiation field, split by a
half-silvered mirror. Then one obtains \cite{AFLP66}
\begin{equation}
|\alpha\rangle_{A} = 
|\mu\alpha\rangle_{B} \otimes |\nu\alpha\rangle_{C} ,
\hspace{1cm} |\mu|^2 + |\nu|^2 = 1 .
\end{equation}
The reason for that is the principal difference in the
structure of the nilpotent Weyl-Heisenberg group $W$ and 
a semisimple group $G$.

We can also prove that the generalized CS are the only states which
factorize upon splitting. We first note that the states of the 
orthonormal basis are obtained by applying the raising operators
to the lowest-weight state one or more times:
\begin{equation}
(E_{\alpha})^p |\Lambda,-\Lambda\rangle = 
|\Lambda,-\Lambda+p\alpha\rangle \times {\rm factor} ,
\end{equation}
where $\alpha > 0$ and $p \in {\Bbb N}$.
Therefore, any state $|\Phi\rangle$ in the Hilbert space can be 
obtained by applying a function of the raising operators
to the lowest-weight state:
\begin{equation}
|\Phi\rangle = f(\{E_{\alpha}\}) |\Lambda,-\Lambda\rangle , 
\hspace{1cm} \alpha > 0 .
\end{equation}
For example, for the SU(2) group one has
\begin{equation}
|j,m\rangle = \left( \begin{array}{c} 2j \\ j+m \end{array}
\right)^{1/2} \frac{ (J_{+})^{j+m} }{ (j+m)! } |j,-j\rangle ,
\end{equation}
and $|\Phi\rangle = f(J_{+}) |j,-j\rangle$. If the state 
$|\Phi_{A}\rangle_{A}$ factorizes upon splitting, 
\begin{equation}
|\Phi_{A}\rangle_{A} = |\Phi_{B}\rangle_{B} \otimes
|\Phi_{C}\rangle_{C} ,
\end{equation}
the following functional equation must be satisfied,
\begin{equation}
f_{A}(\{E_{A \alpha}\}) = f_{B}(\{E_{B \alpha}\})
f_{C}(\{E_{C \alpha}\}) 
\end{equation}
(here and in what follows we consider only the raising operators,
i.e., $\alpha > 0$).
Using Eq.\ (\ref{decay:a}) (for the sake of simplicity, we omit
the identity operators), we obtain
\begin{equation}
f_{A}(\{E_{B \alpha}+E_{C \alpha}\}) = f_{B}(\{E_{B \alpha}\})
f_{C}(\{E_{C \alpha}\}) .   \label{feq}
\end{equation}
Using the same method as in Ref.\ \cite{AFLP66}, we can easily 
prove that Eq.\ (\ref{feq}) has the unique solution --- the 
three functions $f_{X}$ ($X=A,B,C$) must be exponentials:
\begin{equation}
f_{X}(\{E_{X \alpha}\}) = f_{X}(0) \exp\left[ \sum_{\alpha>0} 
\tau_{\alpha} E_{X \alpha} \right] , 
\end{equation}
where $f_{X}(0)$ is a normalization factor.
Here $\tau_{\alpha}$ are complex parameters which are
the same for the three systems $A$, $B$, and $C$. 
For example, in the case of the SU(2) group, we find
\begin{equation}
f_{X}(J_{X +}) = (1+|\zeta|^{2})^{-j_{X}}\exp(\zeta J_{X +}) ,  
\end{equation}
where $X=A,B,C$ and $\zeta \in {\Bbb C}$.
[For compact groups, e.g., for SU(2), the representation spaces 
${\cal H}_{\Lambda}$ are finite-dimensional and $\tau_{\alpha}$ 
can acquire any complex value. For noncompact groups the 
representation spaces ${\cal H}_{\Lambda}$ are infinite-dimensional 
and admissible values of the parameters $\tau_{\alpha}$ are 
determined by the normalization condition; e.g., for SU(1,1) the
coherent-state amplitude $\zeta$ is restricted to the unit disk,
$|\zeta|<1$.]
Recalling Eq.\ (\ref{gencs}) [or Eq.\ (\ref{su2cs}) for the SU(2) 
case], we see that the operator-valued function 
$f_{X}(\{E_{X \alpha}\})$ for each of the three systems ($A$, $B$, 
and $C$) is precisely the function that produces the generalized CS. 
[The constant factor $f_{X}(0)$ is recognized as the normalization 
factor ${\cal N}$ of Eq.\ (\ref{nfactor})]. This completes the proof
of uniqueness.

In conclusion, we have proven that, for quantum systems with 
semisimple Lie group symmetries, the generalized CS are the only 
pure states which factorize upon splitting. Therefore, the 
generalized CS are the unique Bell states for these systems, i.e., 
the unique quantum states which, when split, do not give rise to 
any quantum correlations, and so Bell's inequality is never 
violated. Our result shows that the group-theoretic generalization 
of the Glauber CS preserves their main classical attribute ---
fulfillment of Bell's inequality upon splitting.

C.B. gratefully acknowledges the financial help from the Technion
and thanks the Gutwirth family for the Miriam and Aaron Gutwirth
Memorial Fellowship.
A.M. and M.R. were supported by the Fund for Promotion of Research 
at the Technion, by the Technion VPR Fund --- R. and M. Rochlin
Research Fund, and by GIF --- German-Israeli Foundation for 
Research and Development.


\begin{references}
\bibitem{comment1} The radiation field can be split by a 
half-silvered mirror. Such a lossless beam splitter actually
has two input and two output ports. Therefore, when speaking 
about the radiation field splitting, we assume that the vacuum 
enters the second input port.
\bibitem{comment2} However, the Glauber states are not the only 
        maximally coherent states of the radiation field.
\bibitem{comment:discrete} Continuous representations of noncompact
        semisimple groups could be problematic from the point of 
        view of mathematical rigorousness of our results.
        Therefore, we restrict our discussion to compact groups
        and to discrete representations of noncompact groups.
\bibitem{comment3} We will see that this choice of the reference
        state determines the classical properties of the CS.
        This is related to the fact that the coset space $G/H$
        in this case is a symmetric space that can be considered 
        as the phase space of a classical dynamical system.
\bibitem{comment4} For compact groups one could just as well choose 
        the highest-weight state $|\Lambda,\Lambda\rangle$.
\bibitem{comment:sym} The requirement that $B$ and $C$ must have
        the same symmetry as $A$ is of crucial importance.
        For example, if one considers a boson realization of a 
        Lie group $G$, then boson splitting 
        $a_{A}^{\dagger} = \mu a_{B}^{\dagger} + \nu a_{C}^{\dagger}$ 
        ($|\mu|^2 + |\nu|^2 = 1$) cannot describe a situation when 
        the system $A$, with $G$ as the symmetry group, is split 
        into two subsystems $B$ and $C$, which also possess the 
        symmetry of $G$.
\bibitem{comment:spin} The CS cannot be defined for a spin-zero 
        particle, so its decay cannot be described in the 
        framework of our formalism. The case of a spin-half particle, 
        decaying into spin-half and spin-zero particles, is trivial.
        For $j \geq 1$, the lowest and highest spin states 
        $|j,-j\rangle$ and $|j,j\rangle$, that evidently factorize
        upon splitting, are actually particular cases of the 
        SU(2) CS.
\end{references}
\end{document}